\documentclass[twocolumn,showpacs,amsmath,amssymb,superscriptaddress,nofootinbib]{revtex4}

\newcommand{\tens}[1]{{\boldsymbol{#1}}}
\newcommand{\grad}{{\tens{d}}}

\newcommand{\covd}{{\tens{\nabla}}}

\newcommand{\A}[1]{A^{\!(#1)}}
\newcommand{\Ric}{{\mathbf{Ric}}}

\begin{document}
\title{On conformal Killing--Yano tensors for Pleba\'nski--Demia\'nski family of solutions}

\affiliation{Theoretical Physics Institute, University of Alberta, Edmonton,
Alberta, Canada T6G 2G7}

\affiliation{Institute of Theoretical Physics, Charles University, V
Hole\v{s}ovi\v{c}k\'ach 2, Prague, Czech Republic}

\author{David Kubiz\v n\'ak}

\email{kubiznak@phys.ualberta.ca}

\affiliation{Theoretical Physics Institute, University of Alberta, Edmonton,
Alberta, Canada T6G 2G7}

\affiliation{Institute of Theoretical Physics, Charles University, V
Hole\v{s}ovi\v{c}k\'ach 2, Prague, Czech Republic}

\author{Pavel Krtou\v{s}}

\email{Pavel.Krtous@mff.cuni.cz}

\affiliation{Institute of Theoretical Physics, Charles University, V
Hole\v{s}ovi\v{c}k\'ach 2, Prague, Czech Republic}

\date{July 3, 2007}  

\begin{abstract}
We present the explicit expressions for the 
conformal Killing--Yano tensors for the
Pleba\'nski--Demia\'nski family of type D solutions 
in four dimensions. Some physically important
special cases are discussed in more detail.
In particular, it is demonstrated how the conformal 
Killing--Yano tensor becomes the Killing--Yano tensor 
for the solutions without acceleration.
A possible generalization into higher dimensions is 
studied. Whereas the transition from the 
nonaccelerating to accelerating solutions in 
four dimensions is achieved by the conformal 
rescaling of the metric, we show that such a procedure 
is not sufficiently general in higher dimensions---only 
the maximally symmetric spacetimes in `accelerated' 
coordinates are obtained. 
\end{abstract}


\maketitle

\section{Introduction}
\label{sc:intro}
The complete family of type D spacetimes in four dimensions, including the black-hole spacetimes 
like the Kerr metric, the metrics describing the accelerating sources as the C-metric, or the
non-expanding Kundt's class type D solutions, 
can be represented by the general seven-parametric 
metric discovered by Pleba\'nski and Demia\'nski  \cite{PD}
(cf.~also \cite{Debever:1971}). 
Recently, Griffiths and Podolsk\'y 
\cite{GriffithsPodolsky:2005,GriffithsPodolsky:2006a,GriffithsPodolsky:2006b,PodolskyGriffiths:2006,GriffithsPodolsky:2007} 
put this metric into a new form which enabled 
a better physical interpretation of parameters
and simplified a procedure how to derive all special cases. 
Among subclasses of this solution let us mention
the six-parametric family of metrics without acceleration derived 
and studied already by Carter \cite{Carter:1968} and 
later by Pleba\'nski \cite{Pleb}.

It turns out that the elegant form of the Pleba\'nski--Demia\'nski metric not only yields 
the new solutions in 4D (see, e.g. \cite{Klemm:97, Ortin}), but also inspires for its generalizations 
into higher dimensions which 
became popular in connection with string theories and the brane world models 
with large extra dimensions. Recently, Chen, L\"u, and Pope \cite{CLP} were able to cast the 
Carter's subclass of nonaccelerating solutions 
into higher dimensions---thus constructing the general 
Kerr--NUT--(A)dS metrics in all dimensions.
  
One of the most remarkable properties of the Carter's 
subclass of nonaccelerating solutions,
which is also inherited by its higher dimensional generalizations \cite{FroKu, KuFro},
is the existence of hidden symmetries associated with the Killing--Yano tensor \cite{Penrose, Carter1987}.
Indeed, it is this tensor which is responsible for the `miraculous' properties of the Kerr metric, including
the integrability of geodesic motion or the separability of Hamilton--Jacobi and Klein--Gordon equations
\cite{Carter1, Carter:1968}.
Similar results were obtained recently in higher dimensions \cite{PKVK, FKK, KKPV}.

In four dimensions the integrability conditions for the existence of nondegenerate Killing--Yano 
tensor restricts the Petrov type of spacetime to type D (see, e.g., \cite{Rudiger}). However, 
Demia\'nski and Francaviglia \cite{Dem} demonstrated  that from the known type D solutions 
only spacetimes without acceleration of sources actually admit this tensor.

The purpose of the present paper is to show that the general  Pleba\'nski--Demia\'nski metric 
admits the conformal generalization of Killing--Yano tensor.
We also explicitly demonstrate how in the absence of
acceleration this tensor becomes the known Killing--Yano tensor of 
the Carter's metric.
The particular forms of this tensor for the physically important cases 
are presented. 

We also study a generalization of the Pleba\'nski--Demia\'nski class
into higher dimensions. Namely, we try to `accelerate' the higher-dimensional
Kerr--NUT--(A)dS metric in the same way
as it can be achieved in four dimensions---by 
a conformal rescaling of the metric accompanied with
a modification of the metric functions. 
We demonstrate that this ansatz does not work in odd dimensions 
and in even dimension it leads only to the trivial case of maximally symmetric
spacetimes.
However, it allows us to 
identify the conformal Killing--Yano tensor in 
higher-dimensional flat and (A)dS spacetimes 
related to the `accelerated' coordinates.

\section{Conformal Killing--Yano tensors}
In this section we shall briefly describe the conformal Killing--Yano (CKY) tensors and their basic properties.
The CKY tensors were first proposed by Kashiwada and Tachibana \cite{Kashiwada, Tachibana} as
a generalization  of the Killing--Yano (KY) tensors \cite{Yano}. 
Since then both these tensors found wide applications in 
physics related to hidden (super)symmetries,
conserved quantities, symmetry operators, 
or separation of variables (see, e.g., 
\cite{Carter3, Kamran, Kamran:86, Gibbons:88, Gib, Holten:95, Jonghe,%
Jez:02, Jez:05, Jez:06, Charlton, Benn, Cariglia, Kastor}).

The conformal Killing--Yano (CKY) tensor $\tens{k}$ of rank-$p$ in $D$ dimensions is a $p$-form the covariant derivative of which 
has vanishing harmonic part, i.e., it can be split into the antisymmetric and divergence parts:
\begin{equation}\label{CKY}
\nabla_{\!\mu}k_{\alpha_1\dots \alpha_p}=
\nabla_{\![\mu}k_{\alpha_1\dots \alpha_p]}+
\frac{p}{D-p+1}\,g_{\mu[\alpha_1\!}\nabla_{\!|\kappa|}k^{\kappa}_{\ \alpha_2\dots\alpha_p]}.
\end{equation}
This defining equation is invariant under the Hodge duality;
the antisymmetric part transforms into the divergence part and vice versa.
This implies that the dual $*\tens{k}$ is CKY tensor whenever $\tens{k}$ is CKY tensor.

Two special subclasses of CKY tensors are of particular interest: (a) Killing--Yano tensors \cite{Yano} are those with 
zero divergence part in \eqref{CKY} and (b) closed conformal Killing--Yano tensors with vanishing antisymmetric part
in \eqref{CKY}.
These subclasses transform into each other under the Hodge duality.

In what follows, we shall deal mainly with the rank-2 CKY tensors, which are the only nontrivial in four dimensions,
and which obey the equations
\begin{equation}\label{CKY4D}
\nabla_{\!\mu}k_{\alpha\beta}=
\nabla_{\![\mu}k_{\alpha\beta]}+
{2}\, g_{\mu[\alpha} \xi_{\beta]}\;,
\end{equation}
where we have denoted ${\tens{\xi}}$ the divergence of ${\tens{k}}$:
\begin{equation}\label{xi}
\xi_{\alpha}=\frac{1}{D-1}\nabla_{\!\kappa}k^{\kappa}_{\ \alpha}\;.
\end{equation}
Equivalently,  we could use the alternative definition (see, e.g., \cite{Cariglia, Jez:02})
\begin{equation}\label{alternative}
\nabla_{\!(\mu}k_{\alpha)\beta}=g_{\mu\alpha}\xi_{\beta}-\xi_{(\mu}g_{\alpha)\beta}\;.
\end{equation}

It was demonstrated by Jezierski and \L ukasik \cite{Jez:05} that in  an Einstein space of arbitrary 
dimension the vector $\tens{\xi}$, given by \eqref{xi}, 
either vanishes, which means that the CKY tensor $\tens{k}$
is in fact a KY tensor, or it is a Killing vector.
(We shall see below on a particular example of  Pleba\'nski--Demia\'nski metric
that this may hold more generally, in the presence of electromagnetic field.)
It may be also possible to construct other isometries using the (conformal) KY tensor. 
For example for the general (higher-dimensional) 
Kerr--NUT-(A)dS spacetimes it was demonstrated that all the isometries follow from 
the existence of the principal KY tensor 
\cite{KKPF}.

Whereas the opposite is not generally true (see, e.g., \cite{Ferrando:02} for 
a discussion concerning the nonconformal case), the (conformal) KY tensor 
implies the existence of (conformal) Killing tensor given by:\footnote
{
The (conformal) Killing tensor implies the conserved quantity for the (null) geodesic motion
which is of the second order in momenta. Since (with an arbitrary prefactor) the metric is a (conformal) Killing tensor,
any (conformal) Killing tensor is defined up to a term proportional (with an arbitrary prefactor) to the 
metric.
}
\begin{equation}\label{cKT}
Q_{\alpha\beta}=k_{\alpha\kappa}k_{\beta}^{\ \kappa}\;.
\end{equation} 
This tensor satisfies 
\begin{equation}
\nabla_{\!(\alpha} Q_{\beta\gamma)}=g_{(\alpha\beta}Q_{\gamma)}\;,
\end{equation}
where 
\begin{equation}
Q_{\alpha}=\frac{1}{D+2}(2\nabla_{\!\kappa}Q^{\kappa}_{\ \alpha}+\nabla_{\!\alpha}Q^{\kappa}_{\ \kappa})
\end{equation}
for the conformal Killing tensor whereas it vanishes for the Killing tensor.

\section{Pleba\'nski--Demia\'nski metric}
The original form of the Pleba\'nski--Demia\'nski metric \cite{PD} is given by
\begin{equation}\label{metric}
\begin{split}
\tens{g}=\Omega^{2}&\Bigl[-\frac{Q(\grad\tau-p^2\grad\sigma)^2}{r^2+p^2}+\frac{P(\grad\tau+r^2\grad\sigma)^2}{r^2+p^2}\\
&+\frac{r^2+p^2}{P}\,\grad p^2+\frac{r^2+p^2}{Q}\,\grad r^2\Bigr]\;.
\end{split}
\end{equation}
This metric obeys the Einstein--Maxwell equations with the electric and magnetic charges $e$ and $g$ and the cosmological 
constant $\Lambda$ when functions $P=P(p)$ and $Q=Q(r)$ take the particular form
\begin{equation}\label{functions}
\begin{split}
Q&=k\!+\!e^2\!+\!g^2\!-\!2mr\!+\!\epsilon r^2\!-\!2nr^3\!-\!(k\!+\!\Lambda/3)r^4\;,\\
P&=k\!+\!2np\!-\!\epsilon p^2\!+\!2mp^3\!-\!(k\!+\!e^2\!+\!g^2\!+\!\Lambda/3)p^4\!,
\end{split}
\end{equation}
the conformal factor is
\begin{equation}\label{Omega}
\Omega^{\!-1}=1-pr\;,
\end{equation}
and the vector potential reads
\begin{equation}\label{A}
\tens{A} = -\frac{1}{r^2\!+\!p^2}\,
               \Bigl[ e\,r\,\bigl(\grad \tau \!-\! p^2\,\grad\sigma\bigr)
               +\,g\,p\,\bigl(\grad \tau \!+\! r^2\,\grad\sigma\bigr) \Bigr]\;.
\end{equation}

Our claim is that the general Pleba\'nski--Demia\'nski 
metric \eqref{metric} admits
the conformal Killing--Yano tensor:
\begin{equation}\label{k}
\tens{k}=\Omega^{3}\Bigl[p\,\grad r\wedge(\grad\tau-p^2\grad\sigma)+r\,\grad p\wedge(\grad\tau+r^2\grad\sigma)\Bigr]\;.
\end{equation}
Using the GRTensor, one can easily check that the equations \eqref{CKY4D}, or \eqref{alternative}, are satisfied.
An independent proof is given in the Section \ref{sc:CartPleb}. 

The  dual ${\tens h}=*{\tens k}$ is also a conformal Killing--Yano tensor and it reads 
\begin{equation}\label{bconf}
\tens{h}=\Omega^{3}\Bigl[r\,\grad r\wedge(p^2\grad\sigma-\grad\tau)+p\,\grad p\wedge(r^2\grad\sigma+\grad\tau)\Bigr]\;,
\end{equation}
which is equivalent to 
\begin{equation}\label{bform}
\tens{h}=\Omega^{3}\,\grad{\tens{b}}\;,
\end{equation}
where 
\begin{equation}\label{b}
2\tens{b}=(p^2-r^2)\grad\tau+p^2r^2\grad\sigma\;.
\end{equation}

It is interesting to mention that $\tens{k}$ and $\tens{h}$ are the CKY tensors for the metric \eqref{metric} with 
an arbitrary conformal factor $\Omega$ and arbitrary functions $P(p)$, $Q(r)$, i.e., irrespectively of the fact if 
the metric \eqref{metric} solves the Einstein equations or not. We shall return to this remark in the Section V where we shall also see 
that in the absence of acceleration of sources $\tens{k}$ becomes the KY tensor, whereas $\tens{h}$
becomes the closed CKY tensor. 

Although we deal with a more general space than required by the theorem in \cite{Jez:05},
for ${\Omega}$ given by \eqref{Omega} and for arbitrary functions
$P(p)$ and $Q(r)$ both isometries of the spacetime still follow from
$\tens{h}$ and $\tens{k}$ by eq.~\eqref{xi}:
\begin{equation}
\tens{\xi}_{({h})}=\tens{\partial}_{\tau}\;,\qquad
\tens{\xi}_{{(k)}}=\tens{\partial}_{\sigma}\;.
\end{equation}

The conformal Killing tensor given by \eqref{cKT} which corresponds to $\tens{k}$ reads
\begin{equation}\label{Qk}
\begin{split}
\tens{Q}_{(k)}=\Omega^4&\Bigl[\frac{Qp^2(\grad\tau\!-\!p^2\grad\sigma)^2}{r^2+p^2}+\frac{Pr^2(\grad\tau\!+\!r^2\grad\sigma)^2}{r^2+p^2}\\
&+\frac{r^2(r^2\!+\!p^2)}{P}\,\grad p^2-\frac{p^2(r^2\!+\!p^2)}{Q}\,\grad r^2\Bigr]\,.
\end{split}
\end{equation}
It inherits the `universality' of $\tens{k}$, i.e.,
it is a conformal Killing tensor of the metric \eqref{metric} with an arbitrary $\Omega$, and arbitrary $Q(r)$ and $P(p)$. 
In the absence of acceleration $\tens{Q}_{(k)}$ becomes a Killing tensor which generates the 
Carter's constant for a geodesic motion \cite{Carter1}.
The conformal Killing tensor associated with ${\tens{h}}$ is 
\begin{equation}\label{Qh}
\begin{split}
\tens{Q}_{(h)}=\Omega^4&\Bigl[\frac{Qr^2(\grad\tau\!-\!p^2\grad\sigma)^2}{r^2+p^2}+\frac{Pp^2(\grad\tau\!+\!r^2\grad\sigma)^2}{r^2+p^2}\\
&+\frac{p^2(r^2\!+\!p^2)}{P}\,\grad p^2-\frac{r^2(r^2\!+\!p^2)}{Q}\,\grad r^2\Bigr]\,.
\end{split}
\end{equation}
Both tensors are related as
\begin{equation}\label{QhQk}
  \tens{Q}_{(h)} = \tens{Q}_{(k)} + \Omega^2(p^2-r^2)\, \tens{g}\;.
\end{equation}

Following \cite{GriffithsPodolsky:2006b} one can easily perform the transformations of coordinates and parameters
to obtain the complete family of type D spacetimes and the corresponding forms of CKY tensors.
In the next two sections we shall consider two special cases.
 First we deal with the generalized black holes and then we demonstrate 
what happens when the acceleration of sources is removed.

\section{Generalized black holes}
Following \cite{GriffithsPodolsky:2006b} let's  
introduce two new continuous parameters $\alpha$ (the acceleration) and $\omega$ (the `twist') by 
the rescaling
\begin{equation}
p\to \sqrt{\alpha\omega}p,\
r\to \sqrt{\frac{\alpha}{\omega}}r,\
\sigma\to \sqrt{\frac{\omega}{\alpha^3}}\,\sigma,\
\tau\to \sqrt{\frac{\omega}{\alpha}}\,\tau\;,
\end{equation}
and relabel the other parameters as
\begin{equation}
\begin{split}
m&\to\left(\frac{\alpha}{\omega}\right)^{\!{3}/{2}}\!\!\!m\;,\ 
n\to\left(\frac{\alpha}{\omega}\right)^{\!{3}/{2}}\!\!\!n\;,\
e\to\frac{\alpha}{\omega}\,e\;,\\
g&\to\frac{\alpha}{\omega}\,g\;,\ 
\epsilon\to\frac{\alpha}{\omega}\,\epsilon\;,\quad
k\to\alpha^2k\;.
\end{split}
\end{equation}
Then the metric and the vector potential take the form
\begin{equation}\label{scaledmetric}
\begin{split}
\tens{g}=\Omega^2&\Bigl[-\frac{Q(\grad\tau-\omega p^2\grad\sigma)^2}{r^2+\omega^2p^2}+\frac{P(\omega \grad\tau+r^2\grad\sigma)^2}{r^2+\omega^2p^2}\\
&+\frac{r^2+\omega^2p^2}{P}\,\grad p^2+\frac{r^2+\omega^2p^2}{Q}\,\grad r^2\Bigr]\;,
\end{split}
\end{equation}
\begin{equation}
\tens{A} = -\frac{1}{r^2\!+\!\omega^2p^2}\,
               \Bigl[ e\,r\,\bigl(\grad \tau \!-\!\omega p^2\,\grad\sigma\bigr)
               +\,g\,p\,\bigl(\omega\grad \tau \!+\! r^2\,\grad\sigma\bigr) \Bigr]\;,
\end{equation}
with 
\begin{equation}\label{OmegaA}
\Omega^{\!-1}=1-\alpha\, pr\;,
\end{equation}
and 
\begin{equation}
\begin{split}
Q&=\omega^2 k\!+\!e^2\!+\!g^2\!-\!2mr\!+\!\epsilon r^2\!-\!\frac{2\alpha n}{\omega} r^3
   \!-\!\Bigl(\alpha^2 k\!+\!\frac\Lambda3\Bigr)r^4\;,\\
P&=k\!+\!\frac{2n}{\omega}p\!-\!\epsilon p^2\!+\!2\alpha m p^3
   \!-\!\Bigl(\alpha^2(k\!+\!e^2\!+\!g^2)\!+\!\omega^2\frac\Lambda3\Bigr)p^4\;.
\end{split}
\end{equation}
The CKY tensors are (up to trivial constant factors)
\begin{equation}\label{k2}
\Omega^{-3}\tens{k}=\omega p\,\grad r\wedge(\grad\tau\!-\!\omega p^2\grad\sigma)+
r\,\grad p\wedge(\omega\grad\tau\!+\!r^2\grad\sigma)\;,
\end{equation}
and, $\tens{h}=\Omega^{3}\,\grad{\tens{b}}$, with $\Omega$ given in \eqref{OmegaA} and
\begin{equation}\label{b2}
2\tens{b}=(\omega^2p^2-r^2)\grad\tau+\omega p^2r^2\grad\sigma\;.
\end{equation}

Let's consider two special cases. First, we relabel ${\omega=a}$, perform an
additional coordinate transformation
\begin{equation}\label{transf1}
p\to\cos\theta,\quad 
\tau\to \tau-a\phi,\quad
\sigma\to-\phi\;,
\end{equation}
and set 
\begin{equation}\label{resc1}
k=1,\ \epsilon=1-\alpha^2(a^2+e^2+g^2)-\frac{\Lambda}{3}a^2,\ n=-\alpha a m.
\end{equation}
(One parameter---NUT charge---was set to zero and the scaling freedom was used to eliminate the other two.)
We obtained a six-parametric solution which describes the accelerating 
rotating charged black hole with the cosmological constant:
\begin{equation}\label{BH}
\begin{split}
\tens{g}=&\,\,\Omega^2\Bigl\{-\frac{Q}{\Delta}\bigl[\grad\tau-a\sin^2\!\theta \grad\phi\bigr]^2
+\frac{\Delta}{Q}\grad r^2\\
&+\frac{P}{\Delta}\bigl[a\grad \tau-(r^2+a^2)\grad\phi\bigr]^2+\frac{\Delta}{P} \sin^2\!\theta \grad \theta^2\Bigr\}\;,
\end{split}
\end{equation}
where
\begin{gather}
\Omega^{\!-1}=1-\alpha r \cos\theta\;,\quad
\Delta=r^2+a^2\cos^2\theta\;,\notag\\
Q=(a^2\!+\!e^2\!+\!g^2\!-\!2mr\!+\!r^2)(1\!-\!\alpha^2r^2)-\frac{\Lambda}{3}(a^2\!+\!r^2)r^2\;,\notag\\
\frac{P}{\sin^2\!\theta}=1\!-\!2\alpha m \cos\theta
 \!+\!\bigl[\alpha^2(a^2\!+\!e^2\!+\!g^2)\!+\!\frac{\Lambda a^2}{3}\bigr]\cos^2\theta\;.
\label{cup}
\end{gather}
In the brackets in \eqref{BH} we can easily recognize the familiar form of the Kerr solution. The conformal factor and the
modification of metric functions correspond to the acceleration and the cosmological constant.
The CKY tensor $\tens{k}$ takes the form 
\begin{equation}\label{kgen}
\begin{split}
\Omega^{-3}\tens{k}=&\,
a\cos\theta\, \grad r\!\wedge\left[\grad\tau-a\sin^2\!\theta\, \grad \phi\right]\\
&-r\sin\theta\, \grad \theta\wedge\left[a\grad\tau-(r^2+a^2)\,\grad\phi\right]\;,
\end{split}
\end{equation}
where $\Omega$ is given in \eqref{cup}. Except the conformal factor we recovered the
Killing--Yano tensor for the Kerr metric derived by Penrose and Floyd \cite{Penrose}.

The second interesting example is obtained if instead of \eqref{transf1} and \eqref{resc1} we
perform
\begin{equation}\label{transf2}
p\to\frac{l+a\cos\theta}{\omega},\quad 
\tau\to \tau-\frac{(l+a)^2}{a}\,\phi,\quad
\sigma\to-\frac{\omega}{a}\,\phi\;,
\end{equation}
set the acceleration $\alpha=0$, and adjust 
\begin{equation}
\begin{split}
\epsilon=1-({a^2}/{3}&+2l^2)\Lambda\,,\quad 
n=l+\Lambda l(a^2-4l^2)/3\,,\\
&\omega^2k=(1-l^2\Lambda)(a^2-l^2)\,.
\end{split}
\end{equation}
Then we have a nonaccelerated rotating charged black hole with 
the NUT parameter and cosmological constant:
\begin{equation}\label{NUT}
\begin{split}
\tens{g}=&-\frac{Q}{\Delta}\bigl[\grad\tau-(a\sin^2\theta+4l\sin^2\!\frac{\theta}{2})\,\grad\phi\bigr]^2
+\frac{\Delta}{Q}\,\grad r^2\\
&+\frac{P}{\Delta}\bigl[a\grad \tau-(r^2+(a+l)^2\grad\phi\bigr]^2+\frac{\Delta}{P}\sin^2\!\theta\, \grad \theta^2\;,\!\!
\end{split}
\end{equation}
where
\begin{gather}
\Delta=\,r^2+(l+a\cos\theta)^2\;,\notag\\
\frac{P}{\sin^2\!\theta}=\,1+\frac{4\Lambda}{3}al\cos\theta +\frac{\Lambda}{3}a^2\cos^2\!\theta\;\label{co}\\
\begin{split}\\[-1.5ex]
Q&=a^2\!-\!l^2\!+\!e^2\!+\!g^2\!-\!2mr\!+\!r^2\notag\\
   &\quad-\frac\Lambda3\Bigl(3(a^2\!-\!l^2)\,l^2+({a^2}\!+\!6l^2)\,r^2\!+\!{r^4}\Bigr)\;.\notag
\end{split}
\end{gather}
The CKY tensor $\tens{k}$ becomes the KY tensor (see also the next section) 
and takes the form 
\begin{equation}\label{kgen2}
\begin{split}
\tens{k}=&
(l+a\cos\theta)\,\grad r\!\wedge\left\{\grad \tau+\grad \phi\left[2l(\cos\theta\!-\!1)-
a\sin^2\theta\right]\right\}\\
&-r\sin\theta\,\grad \theta\wedge\left\{a\grad\tau-\grad\phi\left[(l+a)^2+r^2\right]\right\}.
\end{split}\raisetag{2.8ex}
\end{equation}
The dual CKY tensor becomes closed, ${\tens{h}=\grad \tens{b}}$, with 
\begin{equation}
\begin{split}
2\tens{b}&= \bigl((l\!+\!a\cos\theta)^2\!-\!r^2\bigr)\bigl(a\grad\tau\!-\!(l\!+\!a)^2\grad\phi\bigr)\\
   &\quad- r^2(l\!+\!a\cos\theta)^2\grad\phi\;.
\end{split}
\end{equation}

In particular, in vacuum ($e=g=\Lambda=0$) we recover the KY tensor for the Kerr metric ($l=0$), respectively
for the NUT solution ($a=0$) studied recently in \cite{Jez:05}, respectively \cite{Jez:06}.

\section{Carter's metric}
\label{sc:CartPleb}
Let us take the Pleba\'nski--Demia\'nski metric in the form \eqref{scaledmetric} and 
set the acceleration $\alpha=0$, and $\omega=1$. Then the conformal factor becomes $\Omega=1$ and 
we recover the Carter's family of nonaccelerating solutions \cite{Carter:1968}
in the form used in \cite{Pleb}:
\begin{equation}\label{Plebanski}
\begin{split}
\tens{g}=&\ -\frac{Q(\grad\tau- p^2\grad\sigma)^2}{r^2+p^2}+\frac{P( \grad\tau+r^2\grad\sigma)^2}{r^2+p^2}\\
&+\frac{r^2+p^2}{P}\,\grad p^2+\frac{r^2+p^2}{Q}\,\grad r^2\;,
\end{split}
\end{equation}
where
\begin{equation}
\begin{split}
Q=&\,k+e^2+g^2-2mr+\epsilon r^2-\frac{\Lambda}{3}r^4\;,\\
P=&\,k+{2np}-\epsilon p^2-\frac{\Lambda}{3}p^4\;,
\end{split}
\end{equation}
and the vector potential is given again by \eqref{A}.

We also get 
\begin{equation}\label{KY}
\tens{k}=p\,\grad r\wedge(\grad\tau-p^2\grad\sigma)+r\,\grad p\wedge(\grad\tau+r^2\grad\sigma)\;,
\end{equation}
which is the Killing--Yano tensor given by Carter in \cite{Carter1987}. Its dual,
\begin{equation}
\tens{h}=*{\tens{k}}=\grad{\tens{b}}\;,
\end{equation}
with $\tens{b}$ is given by \eqref{b}, becomes the closed CKY tensor.
Again, these properties are independent of the particular form of $P(p)$ and $Q(r)$.
The conformal Killing tensor \eqref{Qk} becomes the Killing tensor
\begin{equation}\label{Qk2}
\begin{split}
\tens{K}=&\,\frac{Qp^2(\grad\tau-p^2\grad\sigma)^2}{r^2+p^2}+\frac{Pr^2(\grad\tau+r^2\grad\sigma)^2}{r^2+p^2}\\
&+\frac{r^2(r^2+p^2)}{P}\,\grad p^2-\frac{p^2(r^2+p^2)}{Q}\,\grad r^2\;.
\end{split}
\end{equation}

Both isometries of spacetime may be derived from 
the existence of $\tens{k}$, but in a different manner than before.
We have $\tens{\xi}_{({h})}=\tens{\partial}_{\tau}$ whereas $\tens{\xi}_{{(k)}}=0$ since $\tens{k}$
is now a KY tensor. Nevertheless, the second isometry is given by 
\begin{equation}
\partial_{\sigma}^{\alpha}= K^{\alpha}_{\ \beta}\,\xi^{\beta}_{(h)}\;.
\end{equation} 

Let us observe that the full Pleba\'nski--Demia\'nski  metric with acceleration 
is related to the Carter's metric 
only by a conformal rescaling
and a modification of the metric functions ${P(p)}$ and ${Q(r)}$. It allows us to use
the theorem---proved recently by Jezierski and \L ukasik \cite{Jez:05}
which says that whenever $\tens{k}$ is the CKY tensor for the metric $\tens{g}$ then $\Omega^{3}\tens{k}$ is the
CKY tensor for the conformally rescaled metric $\Omega^{2}\tens{g}$.
This would justify the transition from the known KY tensor \eqref{KY} 
to the CKY tensor \eqref{k}, up to the fact, that 
in the transition from \eqref{metric} to \eqref{Plebanski}
we also need to change functions ${P(p)}$ and ${Q(r)}$.
Fortunately, as mentioned above, the `universality' of $\tens{k}$, i.e., the property that 
\eqref{KY} remains KY tensor for the metric \eqref{Plebanski} with arbitrary function $P(p)$ and $Q(r)$, can be demonstrated.
Indeed, the only nontrivial components of the covariant derivative ${\tens{\nabla}\tens{k}}$, namely
\begin{equation}
\nabla_{\!p}k_{\sigma r}=\nabla_{\!r}k_{p\sigma}=\nabla_{\!\sigma}k_{rp}=r^2+p^2\;,
\end{equation}
are completely independent of the form of $Q(r)$ and $P(p)$.
(Using these derivatives we easily find that ${\tens{k}}$
is the Killing--Yano tensor satisfying ${\nabla_{\!(\alpha}k_{\beta)\gamma}=0}$.)
Therefore one can start with the metric $\tens{g}$ \eqref{Plebanski}, 
with the KY tensor $\tens{k}$ \eqref{KY}, and with arbitrary functions $P(p)$ and
$Q(r)$ so that, after performing  
the conformal scaling $\tens{g}\to \Omega^2\tens{g}$ we obtain the metric \eqref{metric}. The theorem ensures that 
$\Omega^3\tens{k}$ is the universal CKY tensor of the new metric, and in particular of the 
Pleba\'nski--Demia\'nski solution where $\Omega$ is given by \eqref{Omega} and functions $P(p)$ and $Q(r)$ by 
\eqref{functions}.

\section{Some remarks on higher dimensions}
As we mentioned in the Introduction, the Carter's metric \eqref{Plebanski}
(with charges set to zero) has a form of the 
Chen--L\"u--Pope metric \cite{CLP}
describing the higher-dimensional generally rotating NUT--(A)dS black hole.
Indeed, in an even dimension\footnote{%
In the odd dimensional case there is an additional coordinate 
${\psi_n}$ and additional (nondiagonal) terms in the metric \cite{CLP}. 
Since our procedure of a generalization of Carter's class 
to higher dimensions does not work in odd dimensions, 
we do not discuss this case in more detail.}
${D=2n}$ the metric reads 
\begin{equation}\label{HDBH}
\tens{g}=\sum_{\mu=1}^n\;\biggl[\; \frac{U_\mu}{X_\mu}\,{\grad x_{\mu}^{\;\,2}}
  +\, \frac{X_\mu}{U_\mu}\,\Bigl(\,\sum_{k=0}^{n-1} \A{k}_{\mu}\grad\psi_k \Bigr)^{\!2} \;\biggr]
\end{equation}
with quantities ${U_\mu}$ and ${\A{k}_{\mu}}$ given by
\begin{equation}\label{UAmudef}
\begin{gathered}
  U_{\mu}=\prod_{\substack{\nu=1\\\nu\ne\mu}}^{n}(x_{\nu}^2-x_{\mu}^2)\;,\quad
  \A{k}_{\mu}=\mspace{-25mu}\sum_{\substack{\nu_1,\dots,\nu_k=1\\\nu_1<\dots<\nu_k,\;\nu_i\ne\mu}}^n\mspace{-25mu}x^2_{\nu_1}\dots x^2_{\nu_k}
\end{gathered}
\end{equation}
and with the metric functions
\begin{equation}\label{BHXs}
  X_\mu = b_\mu\, x_\mu + \sum_{k=0}^{n}\, c_{k}\, x_\mu^{2k}\;.
\end{equation}
Here, ${x_\mu}$, ${\mu=1,\dots,n}$, correspond to radial\footnote{%
The radial coordinate (and some related quantities)
are rescaled by the imaginary unit ${i}$ 
in order to put the metric to a more symmetric form (cf., e.g., \cite{CLP}).} 
and latitudinal directions, while ${\psi_j}$, ${j=0,\dots,n-1}$,
to temporal and longitudinal directions. 
The metric can be rewritten in the diagonal form
using a properly chosen orthonormal frame ${\tens{e}^a}$, ${a=1,\dots,D}$ \cite{KKPF}.
The curvature tensors have been computed explicitly in \cite{HamamotoEtal:2007}
and it turns out that the Ricci tensor ${\Ric}$ is diagonal in the frame ${\tens{e}^a}$ as well.

It has been found in \cite{KuFro} and discussed in \cite{KKPF} that the metric \eqref{HDBH}
possesses the Killing--Yano tensor ${\tens{k}}$ and dual closed conformal 
Killing--Yano tensor ${\tens{h}}$ 
\begin{equation}\label{KYHD}
\tens{k}=*\tens{h}\;,\quad
\tens{h} 
 = \frac12\sum_{\mu=1}^n \;\biggl[ \grad x_\mu^2 \wedge \sum_{j=0}^{n-1}\A{j}_\mu\;\grad\psi_j\biggr]
\end{equation}
independently of the specific form of the metric functions ${X_\mu(x_\mu)}$.

For ${D=4}$ we recover the Carter's metric
without electromagnetic field,\footnote{%
For a discussion of `charging' the higher-dimensional
black hole in a way analogous to the four-dimensional Carter's metric
see \cite{Krtous:2007}.} 
${e=g=0}$, by the identification
\begin{equation}
\begin{gathered}
  \psi_0 = t\;,\quad\psi_1=-\sigma\;,\\
  x_1 = ir\;,\quad   X_1 = Q\;,\quad    b_1 = 2im\;,\\
  x_2 = p\;,\quad    X_2 = P\;,\quad    b_2 = 2n\;,\\ 
c_0 = k\;,\quad c_1 = - \epsilon\;,\quad c_2 = -\Lambda/3\;.
\end{gathered}
\end{equation}

It is natural to ask if it is possible to generalize 
the four-dimensional accelerated Pleba\'nski--Demia\'nski
metric into higher dimensions. An obvious procedure to follow would be 
to start with the metric \eqref{HDBH},
rescale it (in analogy with the four-dimensional case)
by a conformal factor ${\Omega^2}$,
\begin{equation}\label{scaledHDBH}
\tilde{\tens{g}}=\Omega^2\tens{g}\;,
\end{equation}
and adjust the metric functions
${X_\mu(x_\mu)}$ in such a way that the scaled metric would satisfy the Einstein equations. 
Due to the same argument which we used in four dimensions 
such a metric would possess a conformal Killing--Yano tensor ${\tilde{\tens{h}}=\Omega^3\tens{h}}$.

The Ricci tensor ${\tilde{\Ric}}$ of the rescaled metric ${\tilde{\tens{g}}}$
is related to the Ricci tensor of the unscaled metric ${\tens{g}}$
by a well known expression (see, e.g., the appendix in \cite{Wald:book1984}),
which can be written as
\begin{equation}\label{confRic}
\begin{split}
  \tilde{\Ric}=\Ric 
     &+(D-2)\Omega\covd\covd\Omega^{\!-1}\\
     &+\tens{g} \Bigl( \Omega\covd^2\Omega^{\!-1} - (D-1)\Omega^2\bigl(\covd\Omega^{\!-1}\bigr)^2\Bigr)\;.
\end{split}
\end{equation}
Here the `square' of 1-forms is defined using 
the inverse unscaled metric ${\tens{g}^{-1}}$.
We require ${\tilde{\Ric} = -\lambda\,\tilde{\tens{g}}}$ 
with ${\lambda}$ proportional to the cosmological constant.
The Ricci tensor ${\tilde{\Ric}}$ thus must be diagonal in the frame ${\tens{e}^a}$.
The conditions on off-diagonal terms give 
the equations for the conformal factor ${\Omega}$. 

In a generic odd dimension these conditions are too strong---they 
admit only a constant conformal factor ${\Omega}$.
In even dimensions the conditions on off-diagonal terms lead to equations
\begin{equation}\label{OmegaCond}
\begin{gathered}
  \Omega^{-1}{}_{\!\!,\mu\nu} = 
    \frac{x_\nu\, \Omega^{-1}{}_{\!\!,\mu}}{x_\nu^2-x_\mu^2}
    +\frac{x_\mu\,\Omega^{-1}{}_{\!\!,\nu}}{x_\mu^2-x_\nu^2}\;,\\
  0=\frac{x_\mu\,\Omega^{-1}{}_{\!\!,\mu}}{x_\nu^2-x_\mu^2}
  +\frac{x_\nu\,\Omega^{-1}{}_{\!\!,\nu}}{x_\mu^2-x_\nu^2}\;.
\end{gathered}
\end{equation}
It gives the conformal factor depending on two constants ${c}$ and ${a}$,
\begin{equation}\label{OmegaHD}
  \Omega^{\!-1} = c + a\, x_1 \dots x_n\;,
\end{equation}
which is obviously a generalization of the 
four-dimensional factor \eqref{OmegaA} (with ${c=1}$ and ${a=i\alpha}$).

Unfortunately, the conditions for diagonal terms of the
Ricci tensor are in even dimensions ${D>4}$ rather restrictive.
Analyzing first the condition for the scalar curvature and then checking
all diagonal terms one finds
that either\footnote{%
The trivial global scaling was eliminated by setting ${a=1}$
in \eqref{dualsol} and ${c=1}$ in \eqref{trivsol}.}
\begin{equation}\label{dualsol}
  \Omega^{\!-1} = x_1\dots x_2\;,\quad
  X_\mu = \bar b_\mu\, x_\mu^{2n-1} + \sum_{k=0}^{n}\, c_{k}\, x_\mu^{2k}\;,
\end{equation}
with ${\lambda=(D-1)\,c_0}$, or
\begin{equation}\label{trivsol}
  \Omega^{\!-1} = 1+a\, x_1\dots x_2\;,\quad
  X_\mu = \sum_{k=0}^{n}\, c_{k}\, x_\mu^{2k}\;,
\end{equation}
with ${\lambda=(D-1)\bigl((-1)^{n-1}c_n+a^2 c_0\bigr)}$.
The first case is not a new solution: the substitution
\begin{equation}\label{dualsoltrans}
  x_\mu = 1/\bar x_\mu\;,\quad
  \psi_j=\bar\psi_{n-1-j}\;,\quad
  X_\mu = \bar x_\mu^{-n+1}\bar X_\mu
\end{equation}
transforms the rescaled metric ${\tilde{\tens{g}}}$ back to
the form \eqref{HDBH} in `barred' coordinates.
In the second case the metric functions ${X_\mu}$
depend on a smaller number of parameters and one has 
to expect that the metric describes only a subclass of the `accelerated black hole solutions'. 
It is actually the trivial subclass---%
it was shown in \cite{HamamotoEtal:2007} that
the metric \eqref{HDBH} with ${X_\mu}$ given by \eqref{trivsol}
represents the maximally symmetric spacetime;
therefore the scaled metric ${\tilde{\tens{g}}}$, being the Einstein space 
conformally related to the maximally symmetric spacetime,
must describe also the maximally symmetric spacetime.
In analogy with the four-dimensional case 
we expect that the metric \eqref{scaledHDBH} with 
metric functions \eqref{trivsol}
describes the Minkowski or \mbox{(anti-)de~Sitter} space
in some kind of `accelerated' coordinates. However, such an interpretation 
needs further detailed study.

We conclude that we did not find a reasonable non-trivial 
generalization of the accelerated Pleba\'nski--Demia\'nski metric to
higher dimensions. However, the physically trivial case \eqref{trivsol}
allows us to write down the conformal Killing-Yano tensor
for the maximally symmetric spacetimes related to the `accelerating' 
`rotating' coordinates ${x_\mu,\,\psi_j}$.

\vspace*{3ex}
\section{Summary} 
We have explicitly demonstrated that the complete family of type D spacetimes which 
can be derived from the Pleba\'nski--Demia\'nski metric possesses the 
conformal generalization of Killing--Yano tensor. In 
the absence of acceleration of sources, i.e., for a special
subclass of solutions described by Carter and Pleba\'nski, this tensor becomes the known 
Killing--Yano tensor. Several examples were discussed in more detail and specific 
forms of these tensors were given.

The Pleba\'nski--Demia\'nski metric also motivates 
for a generalization into higher dimensions.
Its nonaccelerated subclass, the Carter's metric,
has been already generalized into higher dimensions by 
Chen, L\"u, and Pope \cite{CLP}. However, it seems that 
further generalizations, although almost obvious 
at a first sight, cannot be easily obtained.
For example, the attempts to `naturally' charge these solutions failed so far (see, e.g., \cite{Krtous:2007,CL}).
In the present paper we have demonstrated that also the generalization to accelerated solutions is not straightforward.  
In particular, we have shown that the direct analogue of the Pleba\'nski--Demia\'nski complete family (with acceleration)
in higher dimensions cannot be obtained in a manner similar to the four-dimensional 
case, that is, by a conformal scaling of the Chen--L\"u--Pope metric, possibly with the `natural' change of the metric functions.
The question about the existence of the C-metric in higher dimensions therefore still remains 
open.


\section*{Acknowledgments}
D.K.\ is grateful to the Golden Bell Jar Graduate
Scholarship in Physics at the University of Alberta. 
P.K.\ is supported by the grant GA\v{C}R 202/06/0041
and the Czech Ministry of Education under the project
MSM0021610860.

\end{document}